\documentclass[aip,jcp, reprint, amssymb, amsmath,numerical]{revtex4-1}

\usepackage[english]{babel}
\usepackage{amsmath, amssymb, amsthm}
\usepackage{graphicx}

\usepackage{braket}

\usepackage{listings}

\usepackage{upgreek}

\usepackage{color}

\newcommand{\Eq}[1]{Eq.~(\ref{#1})}
\newcommand{\rb}{({\bf r})}
\newcommand{\ro}{{\bf r}}
\newcommand{\rp}{{\bf r'}}
\newcommand{\abs}[1]{\left|#1\right|}
\newcommand{\drdr}{\,d{\bf r} d{\bf r'}}
\newcommand{\Rb}{{\bf R}_B}
\newcommand{\Ra}{{\bf R}_A}

\begin{document}
\title{A New Efficient Method for Calculation of Frenkel Exciton Parameters in Molecular Aggregates}
\author{Per-Arno Pl\"otz}
\affiliation{Institut f\"{u}r Physik, Universit\"{a}t Rostock, D-18051 Rostock, Germany}

\author{Thomas Niehaus}
\affiliation{Institut I - Theoretische Physik, Universit\"at Regensburg, D-93040 Regensburg, Germany}

\author{Oliver K\"uhn}
\email{oliver.kuehn@uni-rostock.de }
\affiliation{Institut f\"{u}r Physik, Universit\"{a}t Rostock, D-18051 Rostock, Germany}

\begin{abstract}
The Frenkel exciton Hamiltonian is at the heart of many simulations of excitation energy transfer in molecular aggregates. It separates the aggregate into Coulomb-coupled monomers. Here it is shown that the respective parameters, i.e. monomeric excitation energies and Coulomb couplings between transition densities, can be efficiently calculated using time-dependent tight-binding-based density functional theory (TD-DFTB). Specifically, Coulomb couplings are expressed in terms of self-consistently determined Mulliken transition charges. The determination of the sign of the coupling requires an additional super-molecule calculation. The approach is applied to two dimer systems. First, formaldehyde oxime for which a detailed comparison with standard DFT using the B3LYP and the PBE functionals is provided. Second,  the Coulomb coupling is explored in dependence on the intermolecular coordinates for a perylene bisimide dimer. This provides structural evidence for the previously observed biphasic aggregation behavior of this dye.
\end{abstract}

\date{\today}
\maketitle

\section{Introduction}
\label{sec:intro}
Excitation energy transfer (EET) has been continuously attracting interest ever since the early studies of F\"orster on fluorescence depolarization of chromophores in solution. \cite{foerster48:55} F\"orster theory owes its success to the establishment of a simple relation between the rate of EET and experimentally observable monomeric absorption and emission spectra.\cite{may11} It is based on the dipole approximation to the Coulomb interaction of electronic transition densities. This approximation, of course, breaks down once the extension of the transition density is of the order of the distance between the monomers. Among the more prominent examples are the light-harvesting antennae of purple bacteria with their closely packed bacteriochlorophyll ring systems, \cite{pullerits96:381,renger01_137} the nanoscale man-made systems such as molecular aggregates \cite{knoester03_1,kuhn11_47} or extended organic conjugated polymer system.\cite{beljonne09:6583} Close proximity of the chromophores comes along not only with the breakdown of the dipole approximation, but with the need to go beyond a second-order rate description, taking into account delocalization of the excitation state over several monomer units.\cite{chachisviis97_7275,scholes11_763}

The theoretical description of EET rests on the model of the Frenkel exciton Hamiltonian, which starts from the separation of the total system into monomeric subunits coupled by the Coulomb interaction.\cite{davydov64_145}
The Frenkel exciton Hamiltonian can be parameterized in several ways. Foremost, its simple structure facilitates an empirical approach, i.e. fitting to experimental data such as different spectroscopic signals (see, e.g. Ref. \citenum{vulto98_9577}). Besides semiempirical methods,  \cite{fron08:1509, olbrich10_12427} in particular time-dependent density functional theory (TD-DFT) enjoys great popularity \cite{madjet06_17268, fuckel08_074505, ambrosek11_17649, ambrosek12_11451} because it often gives a reasonable compromise between accuracy and numerical efficiency (for a critical overview see Ref.\citenum{liu11_1971}). In general, computational approaches can be divided into two categories: First, the aggregate is being viewed as composed of Coulomb-coupled monomers. Obtaining electronic transitions energies for the monomers and their Coulomb couplings an arbitrary aggregate can be formed. \cite{madjet06_17268, fuckel08_074505, ambrosek12_11451}
Second, the aggregate is treated as a super-molecule, usually restricted to a representative dimer. Based on the splitting between excited states and their dependence on the dimer geometry, coupling parameters can be deduced. \cite{fink08:353}

Several approaches have been developed for the calculation of the Coulomb matrix elements. 
Scholes and coworkers \cite{krueger98_2284} have  proposed the  TDC (transition density cube) method, which coarse-grains the configuration space for integration of the transition densities. This way the interaction is expressed as a sum over pairs of TDCs. An alternative was provided  by Renger and coworkers. \cite{madjet06_17268} Their TrEsp (transition charge from electrostatic potential) method employs atomic partial charges in the summation of the Coulomb interaction. These partial charges are obtained by fitting the electrostatic potential of the monomeric transition densities.  As compared with the TDC method this approach has the advantage that numerically converged results can be obtained using some tens of partial charges only (for applications, see, e.g., Refs.\citenum{Renger:2012gb, zhu08:154905}). Still another approach focusses on the  direct implementation of the calculation of the Coulomb coupling utilizing integration and pre-screening tools available in quantum chemistry packages.\cite{fuckel08_074505}

Here we propose and test a new and efficient method
for calculation of Coulomb couplings that is based on 
the time-dependent density functional based tight-binding method
(TD-DFTB).\cite{niehaus01_085108} This scheme may be seen as an
approximate version of TD-DFT, in which computational savings are obtained
by the use of a minimal but accurate atomic orbital basis and the
consistent application of approximations for relevant two-electron
integrals.\cite{niehaus09_38,Grimme2013} These simplifications lead to a significant speed-up with
respect to full TD-DFT calculations and allow the treatment of systems
with several hundred atoms. In contrast to semi-empirical methods for the calculation
of electronic excited states, like for example the 
ZINDO\cite{ridley1973ind} method, TD-DFTB does not rely on free or empirical
parameters which ensures a higher transferability. So far the method
has been applied to such different systems as organic molecules,\cite{Dominguez2013}
semiconductor nanoparticles,\cite{Wang2007,LiZLNF2008} amorphous chalcogenides,\cite{Simdyankin2005a} and biological chromophores.\cite{Wanko2004} The accuracy
achieved is close to the one of the parental TD-DFT method, which also
means that known deficiencies, like the erroneous description of
charge-transfer excited states, are inherited from the latter.           

The paper is organized as follows: In Section~\ref{sec:theory} we give a brief account on the Frenkel exciton Hamiltonian as well as on TD-DFTB. This will lead to the formulation of intermonomeric Coulomb couplings in terms of Mulliken transition charges of TD-DFTB. Applications will be presented in Section~\ref{sec:appl}, first to the formaldehyde oxime dimer, a simple model for which comparison with regular TD-DFT is feasible, and second to a perylene bisimide dimer. Results are summarized in Section~\ref{sec:res}. In the Appendix we give an alternative derivation of the Coulomb coupling, based on a coupled systems partitioning.

\section{Theoretical Methods}
\label{sec:theory}
\subsection{Frenkel Exciton Hamiltonian}
The Hamiltonian of a molecular aggregate of $N_{\rm agg}$ monomers can be written in terms of intramolecular and  intermolecular parts:
\begin{equation}\label{eq:Hagg}
H_{\rm{agg}}=\sum_mH_m+\frac{1}{2}\sum_{m,n}V_{mn}\,.
\end{equation}
Here, $H_m$ describes the individual monomers while $V_{mn}$ is the interaction between pairs of monomers counted by the indices  $m$ and $n$. They can be separated into  Coulomb  and  exchange contributions.\cite{may11} The latter will be neglected  what is justified for not to small intermolecular distances (see below). This results in the following form for the aggregate Hamiltonian:
\begin{eqnarray}
H_{\rm{agg}}&=&\sum_m\sum_{a}\varepsilon_{ma}\ket{\varphi_{ma}}\bra{\varphi_{ma}}\nonumber\\
&+&\frac{1}{2}\sum_{m,n}\sum_{a,b,c,d}J_{mn}(ab,cd)\ket{\varphi_{ma}\varphi_{nb}}\bra{\varphi_{nc}\varphi_{md}}  .
\end{eqnarray}
Here $\ket{\varphi_{ma}}$ denotes the electronic state $a$ of monomer $m$ and $\varepsilon_{ma}$ is the respective energy. 
Using the generalized molecular charge density with the nuclei being at ${\mathbf R}_{A}$ having charge $Z_{A}$ ($e=1$)
\begin{equation}
	\label{eq:dens}
{\mathcal N}_{ab}({\bf r})=\rho^{(m)}_{ab}({\bf r})-\delta_{ab}\sum_{A\in m}Z_A\delta({\bf r}-{\bf R}_A)
\end{equation}
with the electronic density
\begin{equation}
\rho^{(m)}_{ab}({\bf r})=
  \bra{\varphi_{ma}} \hat{n}( {\bf r})  \ket{\varphi_{mb}} = N_{m} \varphi^*_{ma}({\bf r})\varphi_{mb}({\bf r})
\end{equation}
where 
  $\hat{n}( {\bf r})$ is the one-particle electron density operator and $N_m$ the number of electrons of monomer $m$.
\Eq{eq:dens} allows for a compact notation of the Coulomb coupling 
\begin{equation}
	\label{eq:Jmn}
J_{mn}(ab,cd)=\int{}d{\bf r}d{\bf r}'\, \frac{{\mathcal N}_{ad}({\bf r}) {\mathcal N}_{bc}({\bf r}')}{|{\bf r}-{\bf r}'|}\,.
\end{equation}
Notice that in principle the on-site energies and Coulomb couplings are still operators in the space of nuclear coordinates, which are assumed to be fixed and therefore are not further considered.

In the following we will focus on the most common situation of electronic two-level systems ($a=g,e$) and single excitations described by the states
\begin{equation}
\ket{m}=\ket{\varphi_{me}}\prod_{n\neq m}\ket{\varphi_{ng}} \,.
\end{equation}
This yields the well-known one-exciton Hamiltonian:
\begin{equation}
H=\sum_m(\delta_{mn}E_m+J_{mn})\ket{m}\bra{n}
\end{equation}
where $E_m=\varepsilon_{me}-\varepsilon_{mg}$ is the local electronic transition energy and 
\begin{equation}
J_{mn} \equiv J_{mn}(eg,eg)
\end{equation}
couples transition densities at sites $m$ and $n$.

\subsection{Tight-Binding Based DFT}
The ground state DFTB method is derived from a second-order expansion
of the DFT energy functional around a molecular reference density
$\rho_0\rb$.\cite{elstner98_7260} The latter is given by a sum of pseudo-atomic densities
that are obtained from DFT calculations on neutral, spin-unpolarized
atoms. In the basis of the corresponding atomic
orbitals (AO), here termed $\phi_\mu\rb$ (the compound index $\mu=\{Alm\}$ indicates the atom on which the AO is centered, the angular momentum and magnetic quantum number),  the DFTB Hamiltonian reads
\begin{equation}\label{dftb_ham}
    H_{\mu\nu} = H_{\mu\nu}^0+\frac{1}{2} S_{\mu\nu} \sum_{C}(\gamma_{AC} + \gamma_{BC})\Delta q_{C}\text{, } \mu  \in A, \nu \in B.
\end{equation}
The term $H_{\mu\nu}^0$ is the DFT Hamiltonian evaluated at the reference density $\rho_0\rb$ in a two-center approximation, $S_{\mu\nu}$ is the overlap matrix, and $\Delta q_{A}$ denotes the net Mulliken charge on atom A. These charges are obtained in a
self-consistent fashion from the single-particle states $\psi_i =
\sum_\mu c_{\mu i} \phi_\mu$ with the molecular orbital (MO) coefficients $c_{\mu i}$, that are  solutions to the Kohn-Sham equations:
\begin{equation}
\sum_{\nu} \left(H_{\mu\nu} c_{\nu i} -\epsilon_{i}  S_{\mu\nu} c_{\nu i}\right)= 0.
\end{equation}
Here and in the following, we choose the Kohn-Sham states to be
  real quantities.

The term $\gamma_{AB}$ in \Eq{dftb_ham} is a measure of the electron-electron interaction and given by the following two-electron integral:

\begin{gather}  \label{eq:gamma} 
\gamma_{AB} = \iint \drdr \Phi_A(\ro)\left(  \frac{1}{\abs{\ro-\rp}} + f_{\rm xc}[\rho_0](\ro,\rp)  \right) \Phi_B(\rp), \\  \text{with} \qquad \Phi_A(\ro) = \frac{1}{N_A} \sum_{\mu \in A} \left|\phi_\mu(\ro)\right|^2 ,
\end{gather} 
and involves the functional derivative $f_{\rm xc}$ of the DFT exchange-correlation potential. The quantity $N_A$ is the number of basis functions on atom A. Interpolating between the onsite values ($U_A=\gamma_{AA}$) and a pure Coulomb decay at large distance between atoms at $\Ra$ and $\Rb$, the two-electron integral in \Eq{eq:gamma} is conveniently expressed as  $\gamma_{AB}(U_A,U_B,|\Ra-\Rb|)$.\cite{elstner98_7260} The validation of the applied approximations in the DFTB method\cite{bieger85_751,seifert86_267,porezag95_12947,elstner98_7260} and detailed reviews can be found elsewhere.\cite{Frauenheim:2002tx,Seifert:2007bk}

Based on the ground state DFTB method, excited state properties are
accessible by a linear response formulation in the spirit of the
Casida approach in TD-DFT.\cite{casida95_155} In this formalism,
singlet ground to excited state energies $\omega_{eg}$
are obtained from a hermitian eigenvalue problem of dimension N$_\text{occ}$
 $\times$ N$_\text{virt}$, where N$_\text{occ}$ denotes the number of
occupied Kohn-Sham orbitals (labeled by \{$i,j,...$\}) and N$_\text{virt}$ the number of virtual
orbitals (labeled by \{$s,t,...$\}):

 \begin{equation}
  \label{eq_cseigen} \sum_{jt}  \left[     \omega^2_{is} \delta_{ij} \delta_{st} +
     4\sqrt{\omega_{is} } K_{is,jt}
      \sqrt { \omega_{jt}}
\right]
     \; F^{eg}_{jt} =
\omega_{eg}^2 \; F^{eg}_{is} \text{,}
\end{equation}
with $\omega_{is}= \epsilon_s-\epsilon_i$.
Here we have used the fact that the singlet and triplet excitation
manifolds may be disentangled
for systems with a closed shell singlet ground state that we consider in the
following. In the TD-DFTB method,\cite{niehaus01_085108} the so-called
coupling matrix $K_{is,jt}$ is subjected to further
simplifications and finally reads:
\begin{equation}
  \label{dftbsing}
   K_{is,jt} = \sum_{AB} q^{is}_A \gamma_{AB} q^{jt}_B.
\end{equation}
The Mulliken transition charges 
 \begin{equation}
  \label{qij}
  q_A^{is} =
      \frac{1}{2}\sum_{\mu\in A}   \sum_{\nu} \left( c_{\mu i}
      c_{\nu s} S_{\mu\nu} + c_{\nu i} c_{\mu s}S_{\nu\mu}
      \right),
\end{equation}
provide a point charge representation of the Kohn-Sham single-particle
transition densities
$\psi_i\rb \psi_s\rb$   according to
\begin{equation}
  \label{kstradens}
  \psi_i\rb \psi_s\rb \approx \sum_A  q^{is}_A \Phi_A(\ro),
\end{equation}
and can be computed from the ground state MO
coefficients. A detailed discussion of the applied approximation underlying \Eq{eq_cseigen} to \Eq{kstradens} may be found in a recent review on TD-DFTB,\cite{niehaus09_38} which also provides information on the accuracy and limitations of the method. 

Proceeding further, we note that the exact many-body transition density
  is available in TD-DFT and reads \cite{casida95_155,furche01_5982}
  \begin{equation}
    \label{tradns}
    \rho^{(m)}_{eg}({\bf r}) = \sum_{is} \sqrt{\frac{2 \omega_{is}}{\omega_{eg}}}F^{eg}_{is}  \psi_i\rb \psi_s\rb. 
  \end{equation}
  Defining the corresponding many-body transition charges as:
  \begin{equation}
    \label{mulcoup}
    q^{eg}_A=\sum_{is}
    \sqrt{\frac{2 \omega_{is}}{\omega_{eg}}}F^{eg}_{is} q^{is}_A,  
  \end{equation}
  and using Eq. (\ref{kstradens}), we arrive at the final TD-DFTB expression
  for the Coulomb coupling elements $J_{mn}$:

\begin{equation}\label{eq:trans_trans}
J_{mn}(eg,eg)=\sum_{A\in m}\sum_{B\in n}q^{eg}_A q^{ge}_B\zeta_{AB}(|{\bf R}_A-{\bf R}_B|) 
\end{equation}
where  $q^{ge}_A= q^{eg}_A \in \mathbb{R}$.
Here we defined the function 
\begin{equation}
\label{eq:zeta}
\zeta_{AB}(|{\bf R}_A-{\bf R}_B|) = \iint d {\bf r} d {\bf r}' \frac{\Phi_A({\bf r}) \Phi_B({\bf r}')}{\vert {\bf r}-{\bf r}'\vert} \, 
\end{equation}
which in contrast to \Eq{eq:gamma} does not contain the
exchange-correlation term.  Accordingly, the on-site values
($\tilde{U}_A = \zeta_{AA}$) also differ. 
The Frenkel exciton Hamiltonian with TD-DFTB derived transition energies and couplings is termed \emph{tight-binding Frenkel exciton} (TBFE) Hamiltonian. For an account on the other matrix elements of the Coulomb interaction see Appendix B.

Notice that without further information  the presented method will only give the absolute value of the coupling  since the sign of $F^{eg}_{is}$ is arbitrary.  Therefore, the sign of $J_{mn}$ has to be fixed by other means. Options are a comparison with the dipole-dipole approximation or a super-molecule calculation with identification of the coupled states and their oscillator strengths. In terms of application, e.g., in the context of on-the-fly calculations, it is to be expected, that for a given aggregate configuration the sign will not change. In other words, similar configurations will lead to the same sign what requires to perform a single super-molecule calculation only.

At this point is interesting to compare the present approach with the TrEsp method.\cite{madjet06_17268} In TrEsp the electrostatic potential associated with a transition is approximated by a sum of atomic partial charges.
If $Q^{eg}_{A}$ denotes the $A$th charge at $\mathbf{ R}_A$ that is associated with the transition density $\rho_{eg}^{(m)}$ the Coulomb integral in TrEsp is given by
\begin{equation}
\label{ }
J_{mn} \approx \sum_{A \in m} \sum_{B \in n}\frac{Q^{eg}_{A} Q^{ge}_{B}}{|\mathbf{ R}_{A}-\mathbf{ R}_{B}|} \, .
\end{equation}
Comparing this expression to \Eq{eq:trans_trans} the following points should be noted: (i) In both cases transition charges are introduced. In TrEsp these are derived by fitting to quantum chemical data of, in principle, arbitrary accuracy. In the TBFE case the quality of the transition charges depends on the accuracy of the underlying TD-DFTB method. (ii) In \Eq{eq:trans_trans} the extension of the charge distribution is taking into account via $\zeta_{AB}(|{\bf R}_A-{\bf R}_B|)$. While this naturally appears within the TD-DFTB approach, one could argue that related effects are accounted for in TrEsp by virtue of the fitting procedure. (iii) Having in mind large systems with perhaps complicated electron densities, TD-DFTB may be the method of choice anyway and by using \Eq{eq:trans_trans} one will benefit from a consistent description of monomer electronic properties and intermonomer Coulomb couplings. (iv) The TBFE parameters can be easily determined "on-the-fly" in DFTB trajectory simulations. Using TrEsp such an adjustment of the transition densities would be impossible and one relies on \emph{a priori} determined transition charges in the spirit of molecular mechanics force fields (see, e.g., Ref. \citenum{zhu08:154905}).

Finally, we give the expression for the Coulomb coupling in dipole approximation, \cite{may11} which we will use for reference calculations. It reads
\begin{equation}
	\label{eq:dip}
J^{\rm dip}_{m n} =
\frac{ {\mathbf d}_{m,eg}  \cdot{\mathbf d}_{n,ge}} {| {\mathbf X}_{m n} |^3}
- 3 \frac{
( {\mathbf X}_{m n} \cdot {\mathbf d}_{m,eg})
( {\mathbf X}_{m n} \cdot {\mathbf d}_{n,ge})}{| {\mathbf X}_{m n}^{} |^5} 
\end{equation}
where, ${\mathbf X}_{m n}^{}$ is the distance vector between the considered monomers (e.g.\ their centers of mass) and the ${\mathbf d}_{m,eg}$ are transition dipole moments. 
\subsection{Computational Details} 
TD-DFTB calculations were performed with a modified version\footnote{The ng-dftb code is available on request.} of the
TDDFTB$^+$ program package \cite{doi:10.1021/jp070186p,elstner98_7260,
  elstner01_5149} and  the mio-1-1 Slater-Koster
files. \cite{elstner98_7260} In order to study the effect of the
exchange contribution in $\gamma_{AB}$, \Eq{eq:gamma}, as compared
with  $\zeta_{AB}(R)$, \Eq{eq:zeta}, the on-site parameters
  $\tilde{U}_A$ have been calculated, which do not contain the
effect of the exchange-correlation potential.\footnote{These
    parameters were evaluated with an in-house atomic DFT code using
    the PBE exchange-correlation
    functional\cite{perdew96_3865, perdew97_1396} according to $\tilde{U} = (\phi\phi|\phi\phi)$, where
    $\phi$ is the highest occupied atomic orbital of the respective
    element.\cite{elstner98_7260} The parameters are:  
    $\tilde{U}_{\rm{H}}$ = 15.772 eV, $\tilde{U}_{\rm{C}}$ = 14.113
    eV, $\tilde{U}_{\rm{N}}$ = 17.168 and $\tilde{U}_{\rm{O}}$ =
    20.180 eV.}

Below we will also report on possible aggregation structures of PBI-1 dimers. They were obtained by simulated annealing
molecular dynamics on the basis of the ground state DFTB method using the DFTB+ package \cite{doi:10.1021/jp070186p,elstner98_7260}  and including an
empirical dispersion correction.\cite{elstner01_5149} A trajectory has been
equilibrated at 400 K and subsequently cooled down exponentially to 100 K within 1 ps. This was followed by a  1
ps run at 100 K. The temperature was controlled by a Berendsen thermostat.

Reference DFT calculations were performed with the TURBOMOLE 6.4 program suite\cite{TURBOMOLE, treutler95:346, Bauernschmitt1996454} using the 6-311G** basis set and the PBE\cite{perdew96_3865, perdew97_1396} as well as the  the B3LYP\cite{becke:5648, doi:10.1021/j100096a001} functionals. Specifically, Coulomb ($J_{mn}$) and exchange ($K_{mn}$) couplings have been obtained on the basis of the monomeric electronic densities using a pre-release version of the "intact" module of TURBOMOLE. \cite{intact}

\begin{figure}
\includegraphics[width=0.4\textwidth]{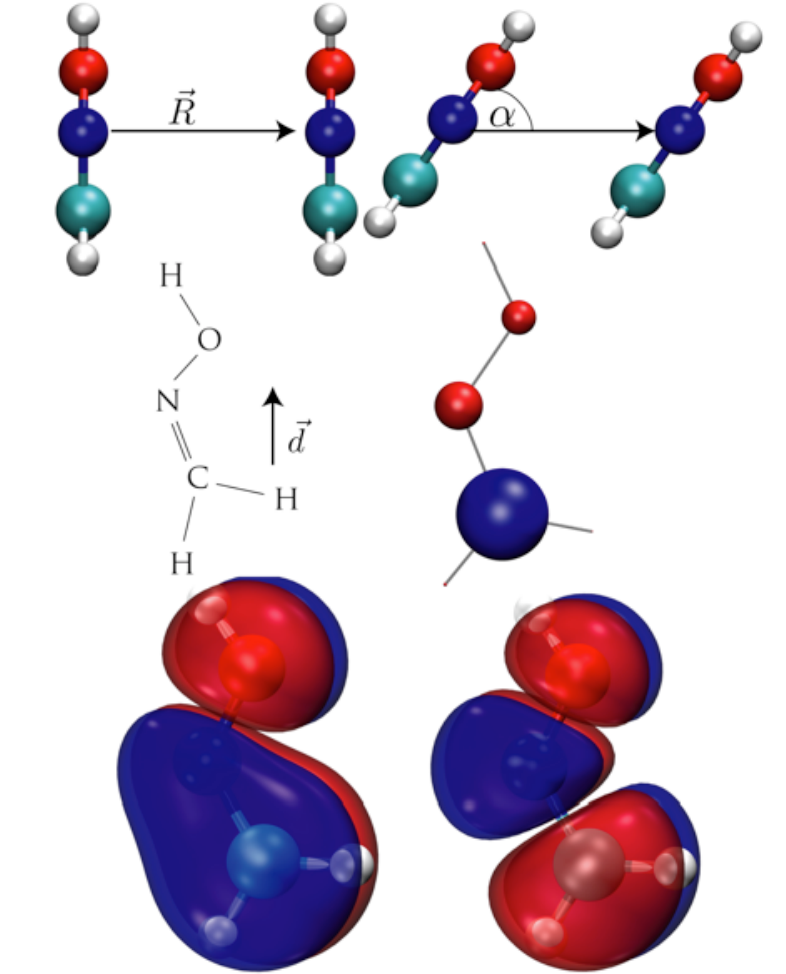} 
\caption{The formaldyhyde oxime dimer (FOD): Upper panel -  definitions of distances and tilt angle; middle panel -  orientation of transition dipole moment and  Mulliken transition charges; lower panel -- TD-DFTB HOMO-1 (left) and  LUMO (right).}
\label{fig:form}
\end{figure}
\begin{center}
\begin{table}[t]
\begin{tabular}{|ccc|ccc|ccc|} 
\hline
\multicolumn{3}{|c|}{B3LYP}  & \multicolumn{3}{c|}{PBE} &\multicolumn{3}{c|}{DFTB} \\ 
exc.& \textit{f}& \textit{E}/eV & exc.& \textit{f} & \textit{E}/eV & exc.& \textit{f} & \textit{E}/eV \\ \hline\hline
12-13 & 0.001 & 5.23 & 12-13& 0.001&5.12&9-10&0.000&5.14\\ \hline
12-14&   0.000 & 6.36 & 12-14& 0.000&5.83 &  &  &\\ \hline
11-14  & 0.000 &  6.79 &  11-14&0.000&6.58  &  &  &\\ \hline
& & & 12-15& 0.054 &7.17 & & &\\ \hline
\textbf{11-13}& \textbf{0.210} &\textbf{7.49}& \textbf{11-13}&\textbf{0.191}& \textbf{7.74} & \textbf{8-10} & \textbf{0.396} &\textbf{7.80}\\ \hline
\end{tabular}
\caption{Assignment of the transitions of the formaldehyde oxime mononer as  obtained by different methods (MOs of leading excitation, oscillator strength, and transition energy, bold-faced the  $\pi\rightarrow\pi^*$ transition, HOMO in DFT (12) and in DFTB (9)).}
\label{tab:zuordnung}
\end{table}
\end{center}

\section{Applications} 
\label{sec:appl}
\subsection{Formaldehyde Oxime Dimer}
The formaldyhyde oxime dimer (FOD), Fig. \ref{fig:form}, has been chosen as a minimal system, which contains O, C, and N atoms typical for chromophores used in molecular aggregates.
To compare the results to DFT-based calculations, the lowest orbitals are qualitatively assigned for the different methods as it is shown in Table \ref{tab:zuordnung}. Our focus will be on the strongest transition, which is of $\pi\rightarrow \pi^*$ (HOMO-1 $\rightarrow$ LUMO) type with the respective single electron orbitals shown in Fig. \ref{fig:form}. DFT with the B3LYP and PBE functionals as well as TD-DFTB give the $\pi\rightarrow \pi^*$ transition between 7.49\,eV and 7.80\,eV. However, the oscillator strength, $f$, is markedly different ranging from 0.191 (PBE) via 0.210 (B3LYP)  to 0.396 (TD-DFTB). This reflects the different transition dipole moments whose absolute values in atomic units are 1.00 (PBE), 1.07 (B3LYP) and 1.44 (TD-DFTB). At around 5.2\,eV all methods predict the essentially dark $n\rightarrow \pi^{*}$ transition. In between these two transitions TD-DFT predicts two (B3LYP), respectively three (PBE) more dark transitions that are not found in TD-DFTB.

Intermolecular configurations for analysis of the Coulomb coupling are scanned along two coordinates (see Fig \ref{fig:form}, upper panel) both starting from a parallel orientation, i.e. the distance $R$ between the molecules (center-of-mass)  and the tilt angle $\alpha$.

\begin{figure}
\includegraphics[width=0.5\textwidth]{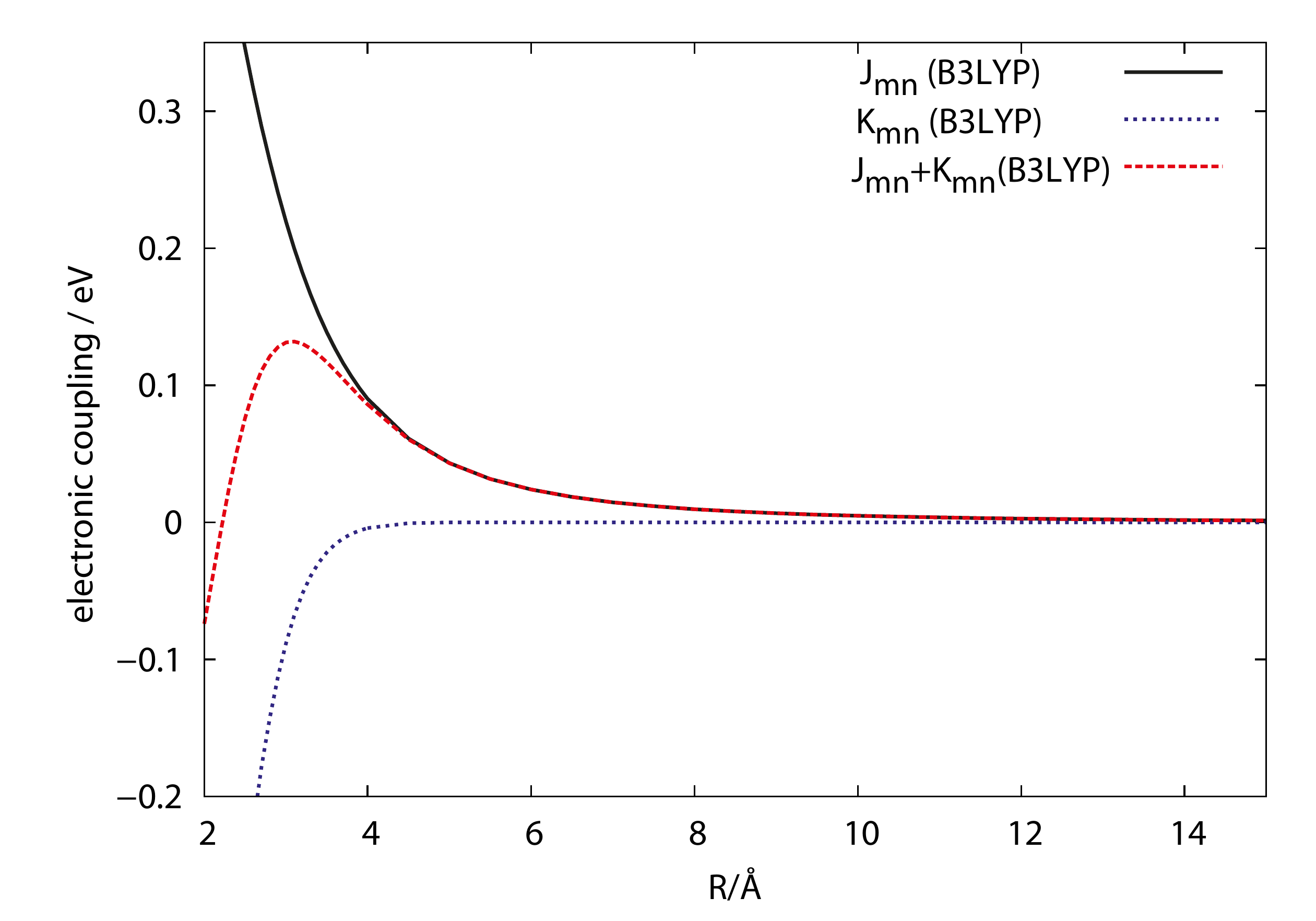}
\caption{Electronic coupling in the FOD for different intermolecular distances at the DFT/B3LYP level of theory. Shown are the Coulomb and exchange contributions as well as the total coupling; for definition of coordinate see Fig. \ref{fig:form}.}
\label{fig:close}
\end{figure}

First we address the parameter range for which the neglect of exchange coupling is justified. In  Fig. \ref{fig:close} Coulomb and exchange contributions as well as their sum is shown for the case of the TD-DFT/B3LYP method. Apparently, the contribution of exchange coupling becomes  negligible for $R>3.5$\,\AA.

\begin{figure}
\includegraphics[width=0.5\textwidth]{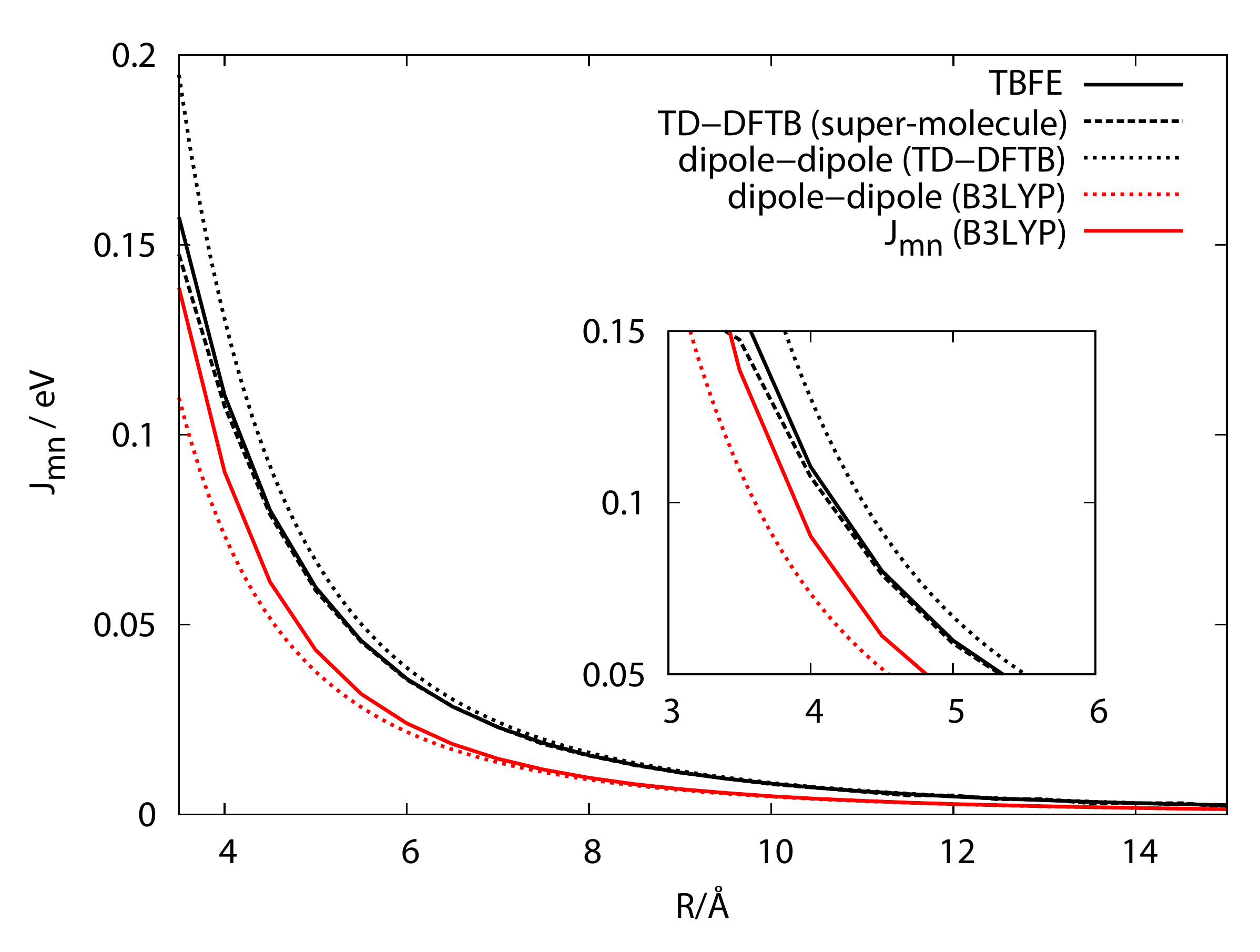}
\caption{Coulomb coupling in FOD as a function of  intermolecular distance and for different methods; for definition of coordinate see Fig. \ref{fig:form}.
}
\label{fig:abstand}
\end{figure}

In Fig. \ref{fig:abstand} a comparison of different methods for calculating the distance-dependent Coulomb coupling is presented and compared to reference TD-DFT/B3LYP calculation of Fig. \ref{fig:close}.  
First, we compare the TBFE and TD-DFTB super-molecule calculations, with the coupling in the latter case  being obtained as half of the energetic separation of those transitions which arise from the local $\pi\rightarrow \pi^*$ transitions.  The distance dependence of the coupling shown in Fig. \ref{fig:abstand} is almost indistinguishable, except at short distances where exchange contributions come into play. Indeed, performing a calculation with 
$\gamma_{AB}$ instead of $\zeta_{AB}$ almost no difference to the super-molecule result had been found (not shown).
Overall, however, the favorable comparison gives indication that the very assumption of coupled two level systems, where the coupling contains only the resonant transition densities, is appropriate on the level of TD-DFTB.
Next we compare the TBFE results with DFT/B3LYP calculations. Here, the TBFE coupling is larger than the DFT/B3LYP one for all distances, with the difference being smaller for short distances. Given the fact that the TD-DFTB transition dipole exceed the one of DFT/B3LYP by about 35\% this is not surprising. However, one should note that the dipole approximation, \Eq{eq:dip}, is valid for distances beyond $R=7$\,\AA. For smaller $R$ the DFT/B3LYP and TD-DFTB values are larger, respectively, smaller than the predictions of the dipole approximation as shown in Fig. \ref{fig:abstand}.

\begin{figure}
\includegraphics[width=0.5\textwidth]{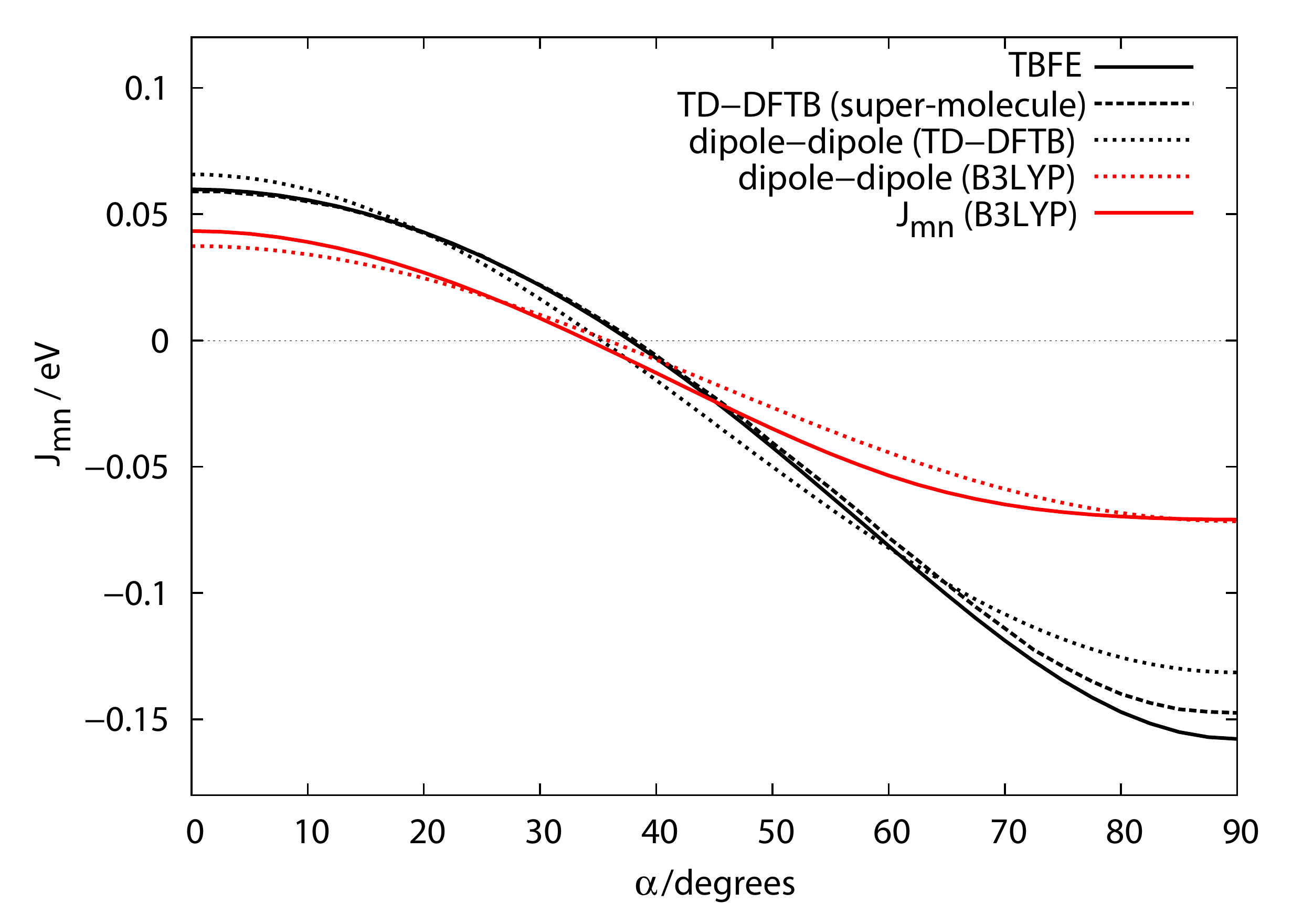}
\caption{Coulomb coupling in FOD  for different tilt angles and methods; for definition of the angle $\alpha$ see Fig. \ref{fig:form}.}
\label{fig:winkel}
\end{figure}

In Fig.~\ref{fig:winkel} results on the dependence of the coupling on the tilt angle are shown.
For the center of mass distance fixed at 5\,\AA{} both molecules are tilted such that the transition dipole moment is slanted towards ${\bf R}$ as indicated in Fig. \ref{fig:form}.  Overall scanning $\alpha$ corresponds to a change from a J-like to an H-like aggregate. TBFE and TD-DFTB super-molecule calculations give almost identical results, supporting the conclusions on the validity of the model drawn before.  Differences with respect to DFT/B3LYP can again be traced to the different transition dipole moments as can be seen by comparison with the dipole approximation.
\subsection{Towards Larger Molecules: Perylene Bisimide Derivative (PBI-1)} 
\begin{figure}
\includegraphics[width=0.4\textwidth]{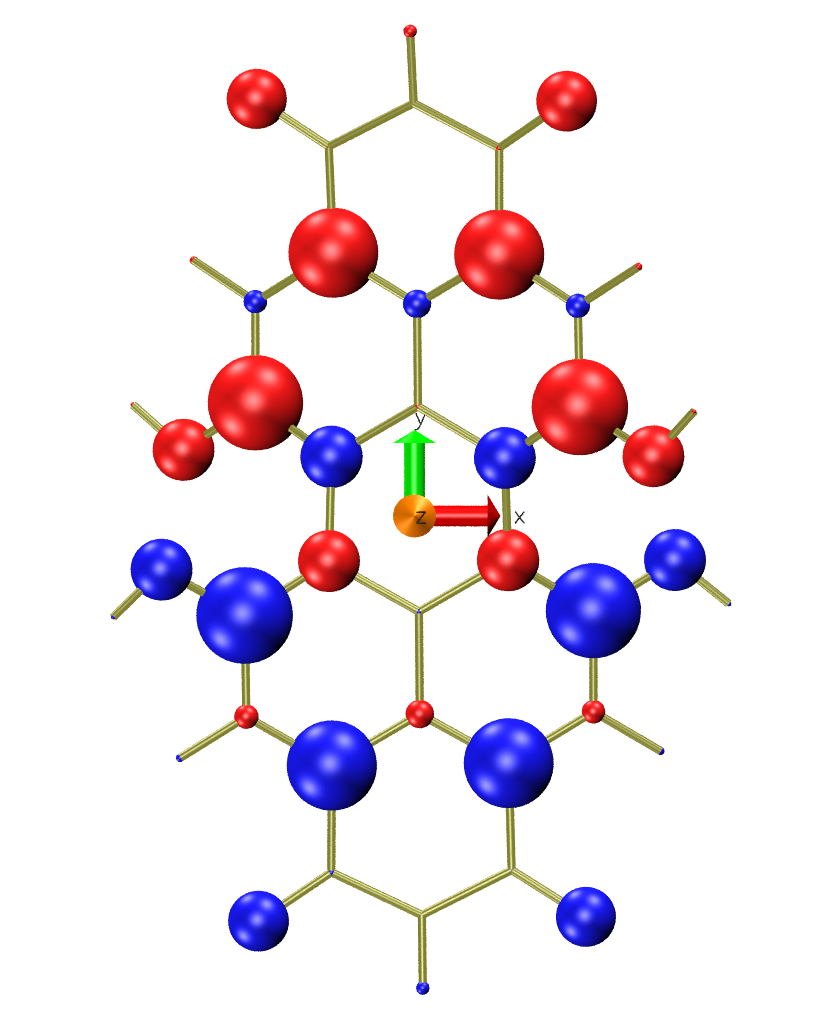} 
\caption{ TD-DFTB Mulliken $\textrm{S}_0\rightarrow\textrm{S}_1$ transition charges for $N$,$N$-Di[$N$-(2-aminoethyl)-3,4,5-tris(dodecyloxy)benzamide]-1,6,7,12-tetra(4-\emph{tert}-butylphenoxy)perylene-3,4:9,10-tetra\-car\-bo\-xy\-lic acid bisimide (PBI-1). Notice that only the (non-planar) perylene core is shown in this figure. The coordinate system for the core is placed at the center of mass with the axes according to the moments of inertia.}
\label{fig:mulliken_pbi}
\end{figure}

For small systems like the FOD there are definitely better methods for obtaining Frenkel exciton parameters. Hence TBFE should be useful for larger systems where highly accurate methods or even standard DFT are not applicable. To illustrate this point, in the following we investigate a dimer comprised of PBI-1 (cf. Fig. \ref{fig:mulliken_pbi}), where each monomer contains 394 atoms. Aggregates of PBI-1 have been synthesized by W\"urthner and coworkers \cite{li08:8074} and the properties of the chromophore were studied, for instance, in Ref. \citenum{ambrosek11_17649}.  Recently, a temperature and concentration dependent study of absorption and fluorescence spectra revealed an unusual behavior.\cite{fennel14_xxx} In particular the fluorescence yield decreases when going from room temperature to about 40\,$^{\circ}$C to increase afterwards. The combination of spectral decomposition and a biphasic kinetic model led to the conclusion that there are two types of dimers, which can be formed in methylcyclohexane solution. One is of J-type and able to form longer aggregates, the other one being of H-type and cannot form longer aggregates due to steric hindrance.

In order to investigate possible aggregation geometries and in particular the behavior of the Coulomb coupling in dependence on intermolecular coordinates we have performed a systematic  TBFE study of the PBI-1 dimer in gas phase. 
The Mulliken transition charges for the perylene core, which comprises the chromophore for the S$_{0}$ $\rightarrow$ S$_{1}$ transition, are shown in Fig. \ref{fig:mulliken_pbi}. The distribution of transition charges reminds on the transition dipole moment, which is oriented along the long ($y$-) axis of the perylene core.

\begin{figure}
\includegraphics[width=0.5\textwidth]{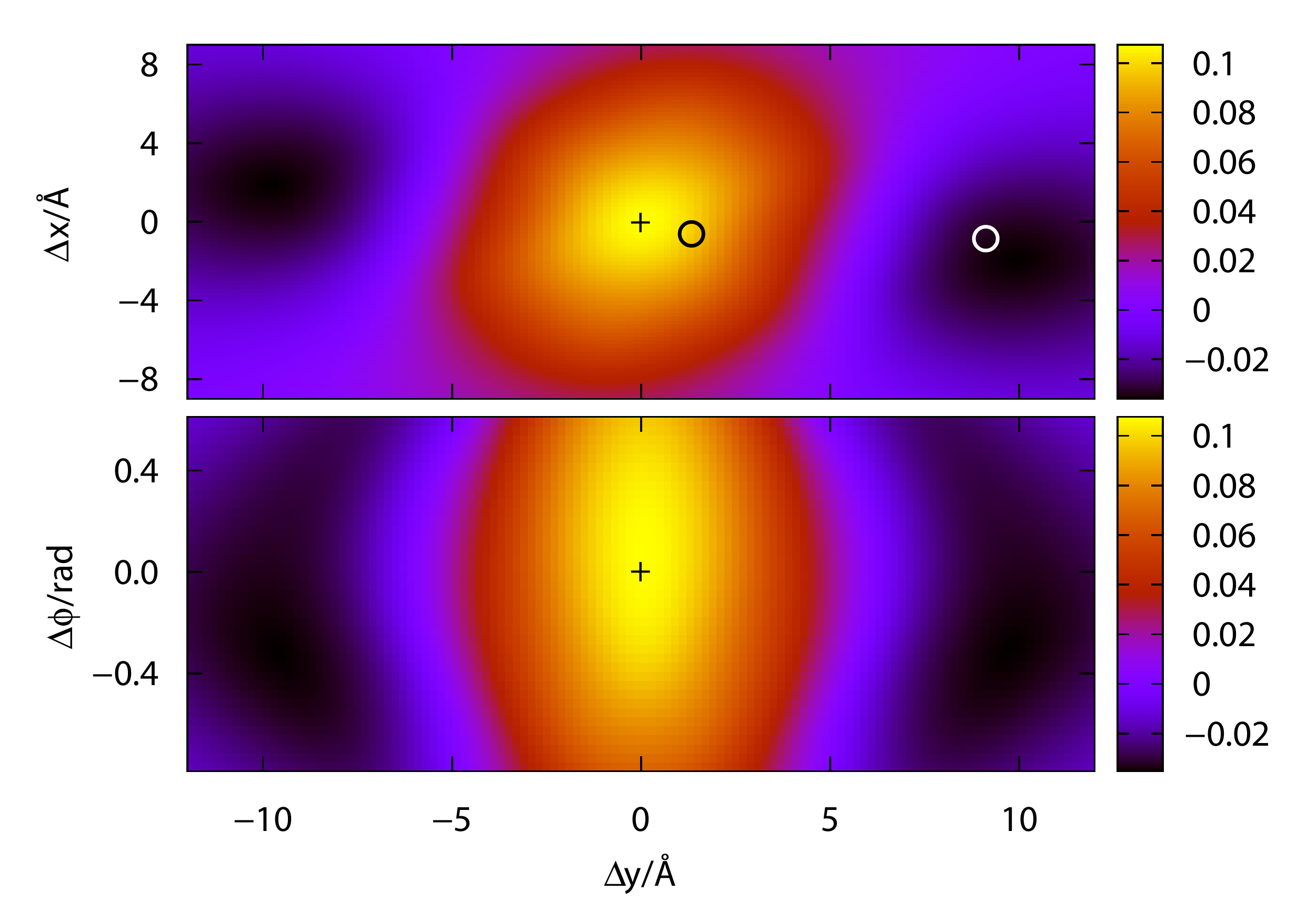}
\caption{TBFE Coulomb couplings in reduced model of PBI-1 dimer for $(x,y)$ shifted geometries (upper panel) and for simultaneous shift along the long axis and rotation around the $z$ axis (lower panel); cf. Fig. \ref{fig:mulliken_pbi}. The cross marks the $(0,0)$ position and the black and white circle the geometries of Fig. \ref{fig:dimere} a and b, respectively.}
\label{fig:scans}
\end{figure}
\begin{figure}
\includegraphics[width=0.5\textwidth]{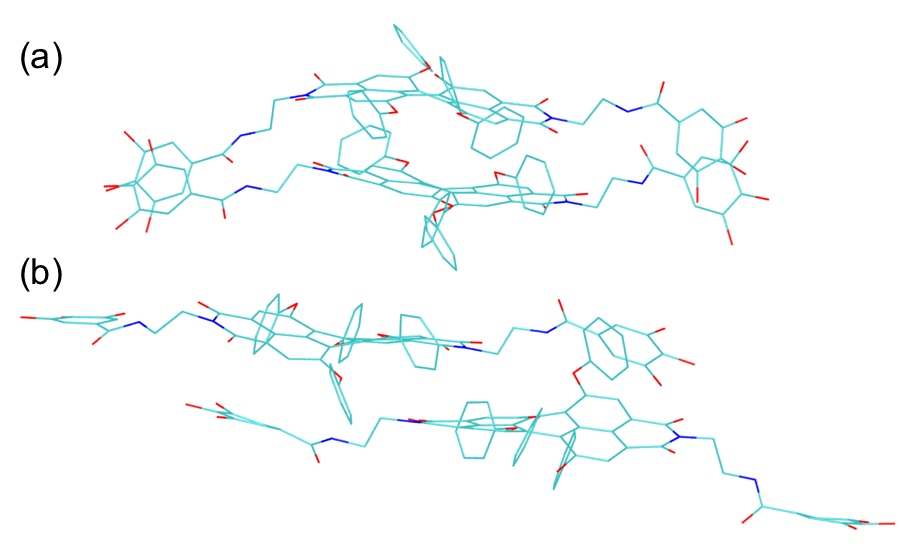} 
\caption{H-type (a) and J-type PBI-1 dimer structure (side groups are not shown for clarity) as obtained from DFTB simulated annealing for the full system (side chains are not shown). Compared to the reference case in Fig. \ref{fig:scans} the H-type structure is shifted by (in \AA) $\Delta x = -0.428$, $\Delta y=  1.341$,
$\Delta z = 0.241$ as well as rotated by about $8^\circ$ with respect to the $z$-axis. For the J-type structure we find 
$\Delta x = -0.663$, $\Delta y=  9.130$, $\Delta z = -0.013$ and rotated by about 44$^\circ$ with respect to the axis (0.21,-0.64,0.74).
}
\label{fig:dimere}
\end{figure}

In the following a dimer formed by two monomers that are reduced to their perylene cores is used to scan the Coulomb coupling as a function of (i) the relative position of the two molecule planes (coordinates $x$ and $y$ in Fig. \ref{fig:mulliken_pbi}) and (ii) the relative position along the long ($y$) axis and the rotation around the $z$ axis by an angle $\phi$. Choosing the coordinate system of Fig. \ref{fig:mulliken_pbi} to be fixed at one monomer, the second monomer is at (0.18,0.08,3.61)\AA, which is used as reference geometry for translations ($\Delta x, \Delta y, \Delta z$). This reference geometry had been obtained from DFT/B3LYP geometry optimization of the PBI-1 dimer capped according to Fig. \ref{fig:dimere}.
For teh present scan at a few points a super-molecule calculation is performed in order to fix the sign of the coupling.

The TBFE scans of the Coulomb coupling are shown in  Fig. \ref{fig:scans}. Apparently, there is a broad region around the origin of axes where the coupling has a positive sign and is of the order $\sim$ 0.08-0.1 eV (upper panel). This holds irrespective of a rotation (lower panel). However, if the two planes are slide with respect to each other along the long axis of the monomer the coupling becomes negative. Negative couplings $\sim$ -0.03 eV are found around $\Delta y = \pm 10$\,\AA~with some flexibility in the rotation angle.

In order to correlate the results of Fig. \ref{fig:scans} to possible structures of the full system we have performed simulated annealing MD simulations. Fig. \ref{fig:dimere} shows two dimer structures obtained from different initial conditions, i.e. configurations around $\Delta x=\Delta y = \Delta z =0$ (panel a) and $\Delta y=-10$\AA{}, $\Delta x = \Delta z =0$ (panel b). The structure in panel (a) having a positive coupling is in stacked H-aggregate configuration. It is marked by a black circle in Fig. \ref{fig:scans} and the actual value for the coupling is $J_{mn}=0.103$ eV; see Tab. \ref{tab:vergleich}. In Ref. \citenum{fennel14_xxx} such type of configuration had been identified as being responsible for the spectral features at intermediate temperatures ($\sim 40^\circ$C). In contrast the structure in panel (b) represents a J-type arrangement (white circle in Fig. \ref{fig:scans}) with the coupling being $J_{mn}=-0.034$ eV; see Tab. \ref{tab:vergleich} and Fig. \ref{fig:scans}. Its slip-stack structure reduces steric hindrance as compared to panel (a) and hence can facilitate the growth of longer aggregates.

\begin{center}
\begin{table}\scalebox{1}{
\begin{tabular}{|l|c|c|}
\hline
configuration& $J_{mn}$/eV (TBFE)& $J_{mn}$/eV (DFT) \\\hline\hline
H &0.091&0.103\\\hline
H +1\,\AA&0.068&0.076\\\hline
J&-0.030&-0.034\\\hline
J+1\,\AA&-0.019&-0.021\\\hline
\end{tabular}}
\caption{Coulomb coupling obtained by TBFE and by DFT/B3LYP for the H- and J-like configurations as shown in Fig. \ref{fig:dimere} and with an additional displacement of 1\,\AA{} along the vector connecting the center of masses of the perylene cores.}.
\label{tab:vergleich}
\end{table}
\end{center}
The reduced PBI-1 model still allows for a comparative DFT/B3LYP
calculation. Such a comparison is performed for selected geometries in
Table \ref{tab:vergleich}. Overall, the agreement is rather good,
deviations range from 10 to 13\%. Notice that in this case the
transitions dipoles of the perylene cores (in atomic units)  are (
-0.57,2.90, 0.80) (TD-DFTB)  and (-0.54,2.66, 0.75) (TD-DFT/B3LYP), i.e. comparable in amplitude and direction. However, at this distance the dipole approximation is not applicable.

\section{Summary} 
\label{sec:res}
In conclusion TD-DFTB offers an efficient method for the calculation
of Frenkel exciton parameters of molecular aggregates. In particular
the Coulomb couplings can be  expressed in terms of Mulliken
transition charges, which are determined within
  the TD-DFTB scheme at low computational cost. This opens the possibility to perform on-the-fly molecular dynamics simulations, with intermolecular forces and Frenkel exciton parameters being determined on the same footing. 

The performance of the new TBFE scheme has been tested for two examples. In case of the FOD it was found that  monomer excitation energies are comparable to DFT/B3LYP/PBE results. The Coulomb coupling reflects the difference in transition dipole strengths if compared to DFT, at least for large distances where the dipole approximation holds. 

In order to demonstrate the applicability to larger systems extensive scans of the Coulomb coupling along intermolecular coordinates of a PBI-1 dimer have been performed. Here, two configurations could be identified for the perylene core system, i.e. a stacked H-like dimer and a slip-stacked J-type dimer. Specific geometries for these dimers were obtained for the full system using DFTB-based simulated annealing. The observation of two different dimers  distinguished by their photophysical properties confirms the biphasic aggregation model introduced in Ref. \citenum{fennel14_xxx}.
\section*{Acknowledgments} 
We gratefully acknowledge financial support by the Deutsche Forschungsgemeinschaft (Sfb 652). We thank Dr. Andreas K\"ohn (Mainz) for providing us with the "intact" module.

\appendix 
\section{Coupled System Formulation}
In the following we present an alternative derivation for the determination of the resonant Coulomb coupling between excitations, which is based on a separation of the Kohn-Sham equation into monomeric and coupling contributions. 
Recalling  \Eq{eq_cseigen} 
 \begin{equation}
  \label{eq_cseigen_II} \sum_{jt}  \left[     \omega^2_{is} \delta_{ij} \delta_{st} +
     4\sqrt{\omega_{is} } K_{is,jt}
      \sqrt { \omega_{jt}}
\right]
     \; F^{eg}_{jt} =
\omega_{eg}^2 \; F^{eg}_{is} \text{,}
\end{equation}
which is adapted to the problem of two subsystems $m$ and $n$ having the same number of electrons. Assuming that the sets of Kohn-Sham orbitals are known for $m$ and $n$ and stay separated (postulates a sufficient large separation and no second-order effects)  $F^{eg}_{is}$ can be sectioned into
\begin{equation}
F^{eg}_{is}=\left\{\begin{array}{cl}F^{eg(mm)}_{is}\,, &\mbox{for }i,s\in{}m\\
F^{eg(mn)}_{is} \,,&\mbox{for }i\in{}m\wedge{}s\in{}n\\
F^{eg(nm)}_{is}\,,&\mbox{for }i\in{}n\wedge{}s\in{}m\\
F^{eg(nn)}_{is}\,, &\mbox{for }i,s\in{}n\,,
\end{array}\right.
\end{equation}
whereas the dimension of $F^{eg}_{is}$ grows by a factor of 4. Introducing $\Gamma$ and $\Delta$ as new variables ($\Gamma,\Delta\in\left\{(mm), (mn), (nm), (nn)\right\}$) Eq. \eqref{eq_cseigen_II} becomes
\begin{eqnarray}
\sum_{jt\Gamma}\left[\omega^2_{is}\delta_{ij}\delta_{st}\delta_{\Gamma\Delta}+4\sqrt{\omega_{is}}K^{\Gamma\Delta}_{is,jt}\sqrt{\omega_{jt}}\right]F^{eg\Gamma}_{jt}&&\nonumber\\
=\omega^2_{eg}F^{eg\Delta}_{is}\,.&&\nonumber\\
\end{eqnarray}
The matrices ${\mathbf K}^{(mm,mm)}$ and ${\mathbf K}^{(nn,nn)}$ stay unchanged, whereby it should be noted that the summation over all atoms has to be performed within the corresponding molecule. ${\mathbf K}^{(mm,nn)}$ and ${\mathbf K}^{(nn,mm)}$ are the couplings between two Kohn-Sham transitions at molecules $m$ and  $n$, which are expressed in terms of Mulliken transition charges as follows
\begin{equation}
K^{(mm,nn)}_{is,jt}=\sum_{A\in{}m}\sum_{B\in{}n}q^{is}_A{}\gamma_{AB}
q^{jt}_B \,.
\end{equation}
Next we recall that only the resonant interaction between transition charges is considered. Hence,  terms like ${\mathbf K}^{(mm,mn)}$ (coupling between a local excitation and a charge transfer transition) and  ${\mathbf K}^{(mn,mn)}$, ${\mathbf K}^{(nm,nm)}$, ${\mathbf K}^{(mn,nm)}$ and ${\mathbf K}^{(nm,mn)}$ (two-electron charge transfer process) are neglected. Along the same lines 
 ${\mathbf F}^{eg(mn)}$ and ${\mathbf F}^{eg(nm)}$ are neglected. This reduces  the dimension of the eigenvalue problem  by a factor 2 and one obtains
\begin{equation}
	\label{eq:CS}
\begin{pmatrix}
{\mathbf M}^{(m)} & {\mathbf V}^{(mn)} \\ {\mathbf V}^{(nm)} & {\mathbf M}^{(n)}
\end{pmatrix}\begin{pmatrix}
{\mathbf F}^{eg(mm)}\\{\mathbf F}^{eg(nn)}
\end{pmatrix}=\omega^2_D\begin{pmatrix}
{\mathbf F}^{eg(mm)}\\{\mathbf F}^{eg(nn)}
\end{pmatrix}
\end{equation}
where ${\mathbf M}^{(m)}$ and ${\mathbf M}^{(n)}$ are the matrices from Eq. \eqref{eq_cseigen_II} for monomers $m$ and $n$, respectively. $ {\mathbf V}^{(mn)}$ describes the interaction and is given by 
\begin{equation}
V^{(mn)}_{is,jt}=4\sqrt{\omega_{is}}K^{(mm,nn)}_{is ,jt}\sqrt{\omega_{jt}}\,.
\end{equation}
To proceed we follow the idea put forward in Ref. \citenum{hsu01_3065}. The transitions for the isolated monomers can be found by solving ${\mathbf M}^{(m)}{\mathbf F}^{eg(mm)}=\omega^2_{eg} {\mathbf F}^{eg(mm)}$ and the equation corresponding to monomer $n$. To calculate the coupling between a pair of resonant transitions (transition energy $\omega_0$ and eigenvector ${\mathbf F}^{(m)}$ and ${\mathbf F}^{(n)}$ as solution for monomer $m$ and $n$, respectively) the interaction will be considered as a perturbation. The solution are two transition frequencies $\omega_+$ and $\omega_-$. The zeroth-order eigenvectors is
\begin{equation}\label{eq:eigen_dimer}
{\mathbf F}^{eg}_{\pm}=\frac{1}{\sqrt{2}}\begin{pmatrix}
{\mathbf F}^{(m)}\\\pm{}{\mathbf F}^{(n)}
\end{pmatrix}
\end{equation} 
with the modified transition frequencies
\begin{equation}
\omega^2_\pm=\frac{\left({\mathbf F}^{(m)\,{\rm T}}, \pm {\mathbf F}^{(n)\,{\rm T}}\right)\begin{pmatrix}
{\mathbf M}^{(m)} & {\mathbf V}^{(mn)} \\ {\mathbf V}^{(nm)} & {\mathbf M}^{(n)}
\end{pmatrix}\begin{pmatrix}
{\mathbf F}^{(m)}\\ \pm{\mathbf F}^{(n)}
\end{pmatrix}}{\left({\mathbf F}^{(m)\,{\rm T}}, \pm {\mathbf F}^{(n)\,{\rm T}}\right)
\begin{pmatrix}
{\mathbf F}^{(m)}\\ \pm{}{\mathbf F}^{(n)}
\end{pmatrix}}\,.
\end{equation}
Since 
\begin{equation}\label{eq:halfgap}
\omega_\pm=\omega_0\pm{}J_{mn}\,,
\end{equation}
it follows that
\begin{eqnarray}
J_{mn}&=&\frac{\omega^2_+-\omega^2_-}{4\omega_0}\nonumber\\
&=&\frac{1}{2\omega_0}{\mathbf F}^{(m)\,{\rm T}}
{\mathbf V}^{(mn)}{\mathbf F}^{(n)}\nonumber\\
&=&\frac{1}{2\omega_0}\sum_{is\in{}m}\sum_{jt\in{}n}F^{(m)}_{is}4\sqrt{\omega_{is}}K^{(mm,nn)}_{is ,jt}\sqrt{\omega_{jt}}F^{(n)}_{jt}\nonumber\\
&=&\sum_{is,A\in{}m}\sum_{jt,B\in{}n}F^{(m)}_{is}\sqrt{\frac{2\omega_{is}}{\omega_0}}q^{is}_A{}
  \gamma_{AB} q^{jt}_B\sqrt{\frac{2\omega_{jt}}{\omega_0}} F^{(n)}_{jt}\,.\nonumber\\
\end{eqnarray}
With Eq. \eqref{mulcoup} the coupling is derived as
\begin{equation}
J_{mn}=\sum_{A\in{}m}\sum_{B\in{}n}q^{eg}_A{}\gamma_{AB}q^{eg}_B\,.
\end{equation}
The definition the coupling as half of the energy gap (Eq. \eqref{eq:halfgap}) includes the exchange interaction. This can be subsequently corrected by replacing $\gamma_{AB}(R)$ by $\zeta_{AB}(R)$ to obtain the previous result Eq. \eqref{eq:trans_trans}.

Thus, compared to the previous expression there is no new result obtained within a coupled systems formulation. The principle assumption in both cases is that the transitions are local. This assumption is either formulated in expressing the transition density by the Mulliken transition charges of a monomer transition (Eq. \eqref{tradns}) or by defining the eigenvector for a transition of the dimer as a combination of the monomer transition vectors (Eq. \eqref{eq:eigen_dimer}). 

Notice that an alternative coupled-systems approach has been proposed by Neugebauer.\cite{neugebauer07:134116} It is based on a frozen-density embedding approach and starts with a different partitioning in \Eq{eq:CS}, which contains the non-additive part of the total kinetic energy in the diagonal blocks (here ${\mathbf M}^{(m)}$). Hence it does not correspond directly to the Frenkel exciton definition.

\section{Other Matrix Elements}

In this appendix it is shown how the other non-resonant Coulomb matrix elements of the one exciton Hamiltonian can be calculated in TD-DFTB. First, we recall that in TD-DFTB the Coulomb coupling is reduced to a description in terms of Mulliken charge fluctuations with respect to a neutral atom reference density. This implies that fluctuating charges contain the full effect of the nuclear charges, i.e. $\mathcal{N}_{ab}({\bf r})$ in \Eq{eq:dens}. For convenience we will use $\Delta q_A \equiv q_A^{gg}$ in the following. Further, excited state Mulliken charges, $q^{ee}_A$ can be obtained using the so-called auxiliary functional approach\cite{furche02_7433}, which has been adopted to TD-DFTB in Ref. \citenum{heringer07_2589}. Hence, the general Coulomb matrix element, \Eq{eq:Jmn}, can be expressed as
  
\begin{equation}
J_{mn}(ab,cd)=\sum_{A\in m}\sum_{B\in n}q^{ad}_A q^{bc}_B\zeta_{AB}(|{\bf R}_A-{\bf R}_B|)\,.
\end{equation}


\end{document}